\documentclass[10pt,conference]{IEEEtran}
\IEEEoverridecommandlockouts
\usepackage{cite}
\usepackage{amsmath,amssymb,amsfonts}
\usepackage{algorithmic}
\usepackage{graphicx}
\usepackage{textcomp}
\usepackage{xcolor,colortbl}
\usepackage{enumitem}
\usepackage{tcolorbox} 
\usepackage{booktabs}
\usepackage{multirow}
\usepackage{url}
\usepackage[T1]{fontenc}
\usepackage{float}
\usepackage[export]{adjustbox}

\definecolor{ForestGreen}{RGB}{34,139,34}
\definecolor{electricindigo}{rgb}{0.44, 0.0, 1.0}
\definecolor{cobalt}{rgb}{0.8, 0.28, 0.8}

\newcommand{\bestmark}[1]{\includegraphics[height=5pt]{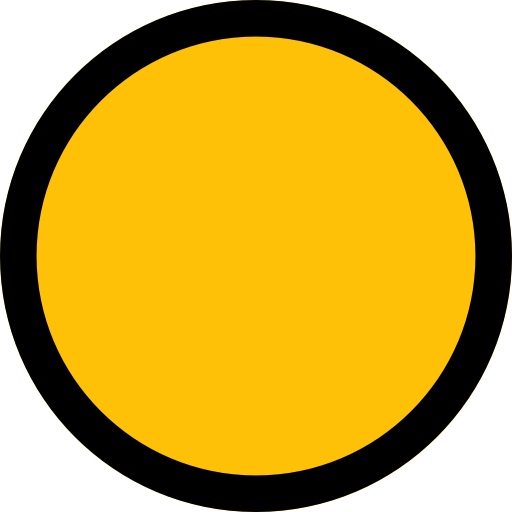} \textbf{#1}}

\begin{document}


\title{T-FREX: A Transformer-based Feature Extraction Method from Mobile App Reviews}









\author{
    Quim Motger\textsuperscript{1}, Alessio Miaschi\textsuperscript{2}, Felice Dell'Orletta\textsuperscript{2}, Xavier Franch\textsuperscript{1}, Jordi Marco\textsuperscript{1}  \\
      \textsuperscript{1}Universitat Polit\`ecnica de Catalunya, Barcelona \\
       {\tt \{quim.motger,xavier.franch,jordi.marco\}@upc.edu}
\\
  \textsuperscript{2}Institute for Computational Linguistics "A. Zampolli" (CNR-ILC), ItaliaNLP Lab, Pisa \\
 {\tt \{alessio.miaschi,felice.dellorletta\}@ilc.cnr.it}
}

\maketitle

\begin{abstract}
Mobile app reviews are a large-scale data source for software-related knowledge generation activities, including software maintenance, evolution and feedback analysis. Effective extraction of features (i.e., functionalities or characteristics) from these reviews is key to support analysis on the acceptance of these features, identification of relevant new feature requests and prioritization of feature development, among others. Traditional methods focus on syntactic pattern-based approaches, typically context-agnostic, evaluated on a closed set of apps, difficult to replicate and limited to a reduced set and domain of apps. Meanwhile, the pervasiveness of Large Language Models (LLMs) based on the Transformer architecture in software engineering tasks lays the groundwork for empirical evaluation of the performance of these models to support feature extraction. In this study, we present T-FREX, a Transformer-based, fully automatic approach for mobile app review feature extraction. 
First, we collect a set of ground truth features from users in a real crowdsourced software recommendation platform and transfer them automatically into a dataset of app reviews. 
Then, we use this newly created dataset to fine-tune multiple LLMs on a named entity recognition task under different data configurations.
We assess the performance of T-FREX with respect to this ground truth, and we complement our analysis by comparing T-FREX with a baseline method from the field. Finally, we assess the quality of new features predicted by T-FREX through an external human evaluation. Results show that T-FREX outperforms on average the traditional syntactic-based method, especially when discovering new features from a domain for which the model has been fine-tuned.
\end{abstract}

\begin{IEEEkeywords}
feature extraction, mobile apps, reviews, token classification, named entity recognition, large language models
\end{IEEEkeywords}

\section{Introduction}
\label{sec:intro}

Mobile app repositories provide valuable access to timely large-scale datasets of software-related information~\cite{Martin2017}. These repositories include heterogeneous, multiple-purpose platforms, from app stores to sideloading repositories and search engines~\cite{ACMNL}. One of the most popular contributions across these platforms is the publication of app reviews, in which users express multiple facets such as personal opinions or experiences, bug reports, inquiries or requests~\cite{Maalej2015}. This information is relevant to multiple software engineering processes, including requirements elicitation and prioritization, release planning, validation analysis and software evolution~\cite{Hassan2022,Yadav2020,Guerrouj2015,Panichella2015,Maalej2015}.

App features are considered a core descriptor for understanding and categorizing app reviews~\cite{Dabrowski2022,Begel2014,Buse2012}. In this context, a feature is considered as a distinct function or capability within a mobile application serving a particular purpose or need~\cite{Dabrowski2023}. App feature extraction supports feature-related knowledge generation, in which mobile app developers can potentially rely on to improve user experience, enhance app functionality, identify user preferences, and make data-driven decisions for app development strategies~\cite{Palomba2015,Guzman2014,Maalej2015}. Consequently, mining large amounts of app reviews to extract app features has become a relevant task.
Nevertheless, mining features from app reviews presents particular challenges. It requires the daily analysis of thousands of short documents, each with a limited length, composed of an average of a few dozen words per review~\cite{Pagano2013}. Beyond measurable characteristics, user-generated documents tend to present multiple informal writing styles and vocabulary, including misspelt words, repetitions or cross-language terminology~\cite{Gao2018}, polarized or biased information, or even noisy and spam content~\cite{Chen2014}.

Consolidated approaches rely on syntactic pattern-matching techniques to retrieve features from app descriptions and reviews~\cite{Dabrowski2023}. 
Nevertheless, several challenges emerge from their applicability, including limited replicability, unavailability of data and a lack of user evaluation~\cite{Dabrowski2023}. Furthermore, rule-based strategies for knowledge generation can be brittle to identify complex patterns, domain-specific terminology, unexpected contents and contextual knowledge, which affects the generalization of these techniques~\cite{Shah2019a}. To overcome these challenges, deep learning strategies, and in particular Large Language Models (LLMs) based on the Transformer architecture~\cite{vaswani2017attention}, have shown promising results in multiple software-related data mining tasks. These approaches leverage the knowledge embedded in these pre-trained models by extending their capabilities through task-specific supervised fine-tuning tasks such as sentiment analysis, 
text classification or named-entity recognition (NER)~\cite{Zhang2020,Ciborowska2022,Yang2022,Tabassum2020}. 

\begin{figure}[htbp]
\centerline{\includegraphics[width=\columnwidth]{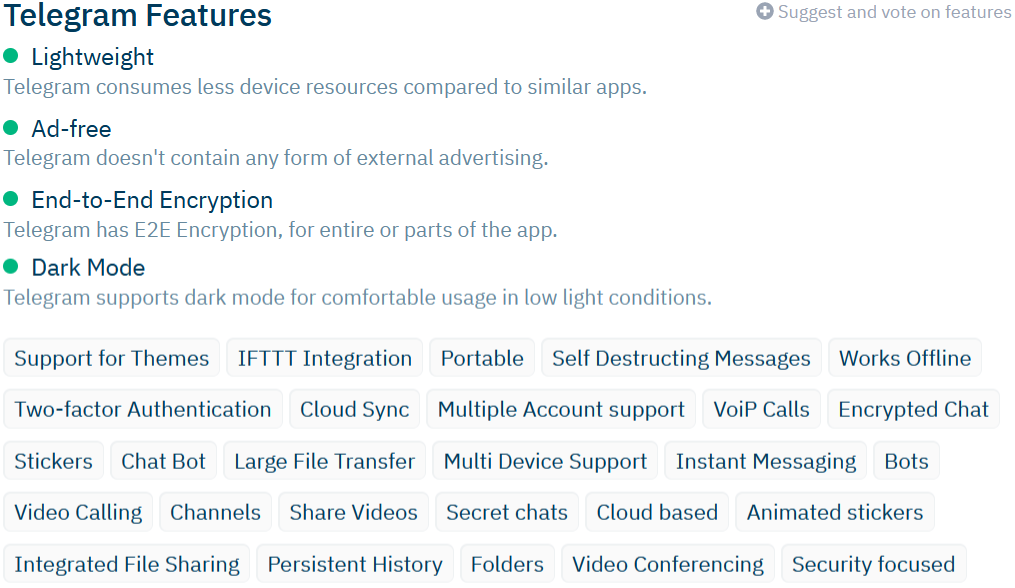}}
\caption{Sample of crowdsourced user annotated features in a software recommendation platform (https://alternativeto.net/) for the Telegram app.}
\label{fig:telegram-features}
\end{figure}

In this paper, we present T-FREX (\textit{Transformer-based FeatuRe EXtraction}), a novel approach to support feature extraction from app reviews using LLMs. Our proposal redefines the app feature extraction problem as a NER task, a specific type of token classification in which tokens referring to a particular entity type (e.g. dates, geopolitical entities, features, etc.) 
are labelled as such. 
Our main contributions are\footnote{GitHub repository: \url{https://github.com/gessi-chatbots/t-frex/}}: (1) a Transformer-based, fully automatic approach for the extraction of mobile app features from user reviews; 
(2) an extensive 
evaluation of the performance of multiple Transformer-based LLMs \cite{vaswani2017attention} 
in different classification scenarios; (3) a reusable fine-tuned 
model 
and 
a ground-truth dataset of annotated app reviews, based on crowdsourced annotated app features extracted from a popular software recommendation platform and automatically transferred into the corpus of app reviews.


\section{Background}
\label{sec:background}


\subsection{Mobile app features}
\label{sec:back-feat-ex}

There are multiple definitions of the term \textit{feature} within related literature according to different dimensions: 
\textit{(i)} scope: definitions refer to functionalities~\cite{Dabrowski2023} (e.g., \textit{send message}), quality aspects~\cite{Jha2019} (e.g., \textit{lightweight}) or both~\cite{kang1990feature}; \textit{(ii)} abstraction: feature expressions vary from generic, high-level categories (e.g., \textit{communication}) to specific, actionable aspects (e.g., \textit{integrated file sharing})~\cite{Guzman2014}; \textit{(iii)} formalization: definitions range from a particular focus on the requirements engineering field, using terms like \textit{logically related system capabilities} and \textit{set of functional requirements}~\cite{Wiegers2013}, to a more user-oriented perspective, referring to a \textit{property}~\cite{Harman2012} or \textit{characteristic}~\cite{KangFeatureOrientedDomain1990} of a mobile app. In the context of grey literature and industrial applications, both functional and quality \textit{features} with different levels of abstraction and formalization are often presented as descriptors at the same hierarchical level. Figure~\ref{fig:telegram-features} illustrates the set of crowdsourced user annotated features (i.e., collaboratively labelled by multiple users) for a given mobile app in a software recommendation platform, for which we find different examples in terms of scope (e.g., \textit{portable} vs. \textit{instant messaging}), abstraction (e.g., \textit{channels} vs. \textit{share videos}) and formalization (e.g., \textit{file sharing} vs. \textit{send file}).

To accommodate our research to both scientific and industrial applications, in this research, we define a \textbf{feature} as a distinct functionality or capability within a mobile application that serves a particular purpose or provides a specific benefit to the user. It is a functional component or attribute of the software designed to perform a well-defined task or address a particular user need, enhancing the utility of the app. 

\subsection{NER using LLMs}
\label{sec:ner-llm}

Our proposal is based on redefining feature extraction as a NER task, one of the most common token classification tasks in the context of natural language understanding (NLU) and for which LLMs have been largely used in the past few years~\cite{jehangir2023survey}. Given a set of app reviews $R$ of size $n$, the tokenized representation of a given $r_i \in R$ is expressed as $T(r_i) = [t_{i1}, t_{i2},...,t_{im}]$, where each $t_{ij} \in T(r_i)$ represents a token from the original review $r_i$. The \textit{feature extraction} task consists in identifying sequences of contiguous tokens $T_{if} \subseteq T(r_i)$ composing the written expression of a feature. Consequently, features are extracted within the context of a particular app review. This context dimension is key, as a particular sequence of tokens $T_f$ might refer to a feature within a given review for a given app, but the same sequence $T_f$ might not refer to a feature in another context. To this end, the dynamic attention mechanism enables LLMs to attend to crucial contextual information within the context of a review from a mobile app of a specific category. For example, in the review sentence ``\textit{I find that managing my channels is quite frustrating}'', the word \textit{channels} refers to an actual feature of a communication mobile app. Contrarily, in the review sentence ``\textit{This art app offers diverse channels for unleashing creativity}'', which belongs to a mobile app from the arts and design category, \textit{channels} is not an actual feature, despite having the same Part-of-Speech (PoS) tag (NOUN) and syntactic dependency role (direct object). Traditional syntactic-based approaches lack the potential to determine how contextual information suits a more fine-grained selection of features.

\begin{figure}[htbp]
\centerline{\includegraphics[width=\columnwidth]{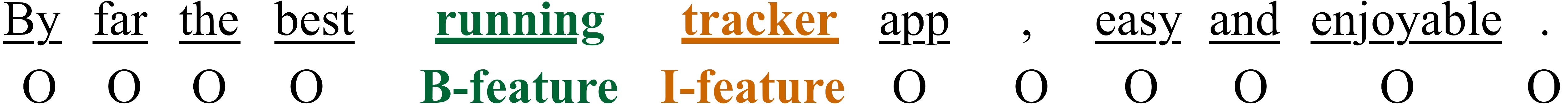}}
\caption{Example of NER task on a mobile app review.}
\label{fig:ner-example}
\end{figure}

In token-based classification tasks, the input can range from a sentence to a paragraph or even an entire document. Our approach is defined at the sentence level, allowing for more granular analysis and facilitating efficient processing.
Therefore, tokenization is refined by splitting the reviews into sentences, $T(r_i) = [s_{i1}, s_{i2},...,s_{ip}]$, where each sentence is a subsequence of the original tokenization, $s_{ik} \subseteq T(r_i)$. 
Given a sentence input $s_{ik}$, a token classification model assigns one unique label to each token, indicating whether the token is the beginning of a feature named entity (\textit{B-feature}), an internally contained element of a feature entity (\textit{I-feature}), or none (\textit{O}). 
Figure~\ref{fig:ner-example} illustrates the NER output on a mobile app review. 
For simplicity, this example is architecture-agnostic, meaning that we consider each word from the original review individually, ignoring special tokens or multiple tokens referring to the same word.
While numerous models are publicly available for generic types of NER (e.g., dates, locations, persons, e-mail addresses...) and for some specific domains (e.g. medical and legal domains) \cite{hakala-pyysalo-2019-biomedical,Tarcar2019HealthcareNM,gu2020named}, to the best of our knowledge there are no proposals for the identification of app feature entities exploiting fine-tuned LLMs.

\section{Research method}
\label{sec:method}


Our research aims to demonstrate the hypothesis that \textbf{Transformer-based LLMs can significantly enhance mobile app feature extraction tasks by redefining this process as a token classification problem}.
Consequently, we conducted a sample study as defined by Stol and Fitzgerald~\cite{Stol2018}, which aims at maximizing the generalization of the feature extraction task over the population of mobile apps publicly available in mobile app repositories. 
To this end, we leverage crowdsourced annotated data from actual mobile app users, using AlternativeTo\footnote{https://alternativeto.net/} 
as the main source for ground-truth knowledge generation, as a relevant representative of search engines in the context of mobile app repositories~\cite{Motger2023}. This platform provides, for each mobile app, a list of features which have been suggested and ranked by real users, while they are being used for navigating the catalogue of mobile applications (as illustrated previously in Figure~\ref{fig:telegram-features}). More details on the data collection and annotation processes are depicted in Section~\ref{sec:data}.

To assess the validity of T-FREX, we guide our research through the following research questions:





\noindent\textbf{RQ1)} What is the effectiveness of T-FREX using different LLMs with respect to crowdsourced user-annotated features?

\noindent\textbf{RQ2)} What is the effectiveness of T-FREX compared to traditional feature extraction methods (i.e., SAFE approach)?

\noindent\textbf{RQ3)} What is the effectiveness of T-FREX with respect to new, undocumented features?

RQ1 is defined to assess the effectiveness of our approach in terms of functional suitability. We present the design and development results of an end-to-end pipeline for fine-tuning and using different types of LLMs under different data configurations. This analysis will allow us to gain a deeper understanding of how different LLM architectures behave under different training datasets. Moreover, RQ1 provides token-level empirical evaluation results for the overall quality of the NER (i.e. token classification) task.

RQ2 is intended to compare T-FREX performance at feature level with respect to a standard baseline method in the field of feature extraction. We selected the SAFE approach as a baseline for a comparative analysis~\cite{Johann2017a}, as it is considered the most consolidated approach for mobile app feature extraction in the software engineering field (see Section~\ref{sec:rel-work} for more details on related work and comparison with other approaches). Given that the original study does not provide a publicly available replication package, we built on the work of Shahe et al.~\cite{Shah2019a}, who conducted a replication study and distributed a replicated development of the SAFE approach.

Finally, RQ3 is designed to analyse how T-FREX generalizes and overcomes the constraints of limited, domain-specific datasets. While the use of data generated and consumed by real users offers multiple advantages, we have no control in the extent and representativeness with respect to the complete set of features exposed by mobile apps. This entails that our model might predict features that are not included in the ground truth (i.e., \textit{false positive}), while this might simply relate to incompleteness of user annotated data. Hence, we propose to overcome this limitation while also gaining deeper insights on generalization of our model by conducting a human evaluation on new features predicted by our model (i.e., features not included in the ground truth dataset).
\section{Design}
\label{sec:design}

\begin{figure*}[htbp]
\centerline{\includegraphics[width=\textwidth]{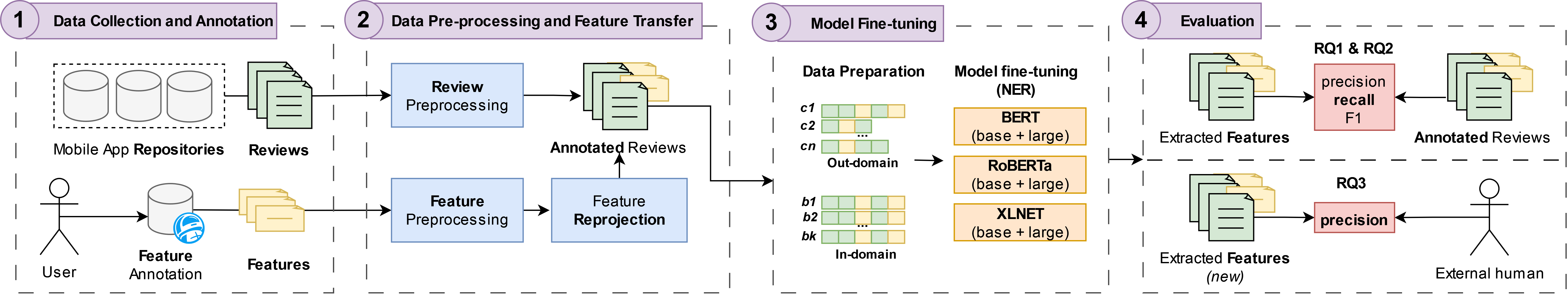}}
\caption{Research design overview.}
\label{fig:design}
\end{figure*}

Figure~\ref{fig:design} shows an overview of our research. We elaborate the details in the upcoming subsections.

\subsection{Data Collection and Annotation}
\label{sec:data}

While there is related work publishing gold datasets with expert feature annotations~\cite{Dabrowski2023}, our approach is intended 
to leverage real user crowdsourced annotation, as well as to assess its generalization and applicability in real, practical uses cases. However, published datasets are typically internally annotated by human coders, very limited in terms of number of applications (e.g., between 8-10 mobile apps) and domains, and focused on productivity and communication apps, excluding more expert, domain-specific categories like navigation, sports or weather~\cite{Johann2017a,Shah2019a,DRAGONI20191103,Dabrowski2023,Guzman2014}. Therefore, we opted to build our own dataset of mobile app reviews with annotated features generated by real users. We built on the work of Motger et al.~\cite{Motger2023}, who collected and published a sample dataset of 639 mobile apps with 622,370 reviews from multiple categories, to which we applied the following extensions:

\begin{itemize}
    \item We extended the mobile app metadata with the official category from Google Play as gold knowledge for the category-based analysis of the feature extraction task. The original dataset included custom defined categories based on a keyword-based search of domain-related terms. Instead, we propose to use the taxonomy of categories defined by Google Play~\cite{GooglePlayCategories}, which is considered as the largest, most relevant mobile app store worldwide~\cite{ACMNL}. 
    \item We limited the mobile apps included in our study to the 10 most frequent Google Play categories in the original dataset in number of mobile apps, excluding those categories with a minimal representation ($\leq 5$ apps) to ensure relevant statistical inference. Exceptionally, we excluded ``GAME" related categories, considered as a special kind of mobile apps with a different \textit{feature} conception~\cite{Dabrowski2023}. 
    \item We extended the annotated features for a given mobile app using AlternativeTo features as ground truth. These features are voted by logged-in users of the software recommendation platform, and they are ranked and sorted by absolute number of votes. To obtain this data, we reused the web scraping data collection mechanisms originally developed by Motger et al.~\cite{Motger2023}.
\end{itemize}

Table~\ref{tab:dataset} summarizes the resulting dataset distributed according to the Google Play category to which the apps belong to. It is important to highlight that multiple distinct features might belong to multiple categories (e.g., \textit{video call} is a feature for both SOCIAL and COMMUNICATION mobile apps). Notice that the last row refers to the total number of feature annotations in the complete corpus of annotated reviews, which we explain in more detail in Section~\ref{sec:prep-repr}.

\subsection{Data Pre-processing and Feature Transfer}
\label{sec:prep-repr}

\begin{table*}[]
\centering
\caption{Distribution of mobile apps, reviews and features in the dataset, sorted by decreasing order by number of distinct features. Category abbreviations refer to: Productivity (PROD.), Communication (COMM.), Personalization (PERS.).
}
\resizebox{\textwidth}{!}{%
\begin{tabular}{lrrrrrrrrrrr}
\hline
\textbf{Metric} &
  \multicolumn{1}{r}{\textbf{PROD.}} &
  \multicolumn{1}{r}{\textbf{COMM.}} &
  \multicolumn{1}{r}{\textbf{TOOLS}} &
  \multicolumn{1}{r}{\textbf{SOCIAL}} &
  \multicolumn{1}{r}{\textbf{HEALTH}} &
  \multicolumn{1}{r}{\textbf{PERS.}} &
  \multicolumn{1}{r}{\textbf{TRAVEL}} &
  \multicolumn{1}{r}{\textbf{MAPS}} &
  \multicolumn{1}{r}{\textbf{LIFESTYLE}} &
  \multicolumn{1}{r}{\textbf{WEATHER}} &
  \multicolumn{1}{r}{\textbf{ALL}} \\ \hline
\textbf{\#apps}               & 137     & 51      & 58     & 14     & 75     & 6     & 19     & 31    & 12    & 65     & \textbf{468}     \\
\textbf{\#reviews}            & 7,348   & 7,003   & 4,321  & 819    & 2,154  & 112   & 530    & 284   & 344   & 901    & \textbf{23,816}  \\
\textbf{\#sentences}          & 8,604   & 8,135   & 5,402  & 899    & 2,330  & 118   & 602    & 315   & 391   & 984    & \textbf{27,780}  \\
\textbf{\#tokens}             & 148,172 & 134,833 & 93,395 & 15,597 & 40,907 & 2,022 & 11,105 & 5,868 & 8,044 & 15,439 & \textbf{475,382} \\
\textbf{\#features \textit{(distinct)}} & 77      & 54      & 50     & 26     & 23     & 19    & 17     & 12    & 10    & 7      & \textbf{198}     \\ 
\textbf{\#features \textit{(annotated)}}           & 9,866   & 9,800   & 6,626  & 1,049  & 2,524  & 127   & 662    & 333   & 419   & 1,037  & \textbf{32,443}  \\ \hline
\end{tabular}%
}
\label{tab:dataset}
\end{table*}

Let $F = \{f_1, f_2, ..., f_q\}$ be the set of crowdsourced feature annotations defined at app entity level. To train and evaluate our model, these features are transferred into the corpus of app reviews $R$. This results in the annotation of all tokens $t \in T(r)$ for each review $r \in R$ with the corresponding name entity label $L = \{\textnormal{\textit{O}}, \textnormal{\textit{B-feature}}, \textnormal{\textit{I-feature}}\}$. 
To this end, 
we used Stanza's neural pipeline~\cite{Stanza} to pre-process and transform both corpus $F$ and $R$ into their correspondent CoNLL-U format representation~\cite{CoNLL}, which includes for each token $t$ a list of syntactic and morphological features. We use this format to facilitate replicability of our approach and reusability of the resulting dataset. Specifically, the pre-processing pipeline included: (\textit{i})~tokenization, (\textit{ii}) multi-word token expansion, (\textit{iii}) PoS tagging, (\textit{iv}) morphological feature extraction, and (\textit{v}) lemmatization.
Feature transfer is then applied to exact matches between the CoNLL representation of a given feature $f$ and a contiguous sequence of tokens of a given review $r$ so that $f \subseteq T(r)$ after pre-processing $r$ and $f$. Given that not all features $f \in F$ relate to actual features in different contexts, we scoped this label transfer to features originally extracted from the same application to which the review belonged to. This means that, for ground truth generation, the example used in Section~\ref{sec:ner-llm} for context-dependent features (i.e., use of \textit{channels} as a feature from communication apps) is not considered as an actual feature in the context of a different app (i.e., the arts and design app). As a result, the CoNLL-U representation of $R$ is extended with an additional annotation for each token $t$, represented by one of the labels in $L = \{\textnormal{\textit{O}}, \textnormal{\textit{B-feature}}, \textnormal{\textit{I-feature}}\}$.

Table~\ref{tab:dataset} reports the total amount of feature annotations transferred into the corpus of reviews $R$ (last row).



\subsection{Model Fine-tuning}

The pre-processed and annotated corpus $R$ serves as the primary input for training various LLMs under diverse data configurations. In this section, we elaborate on the reasoning behind our choices regarding the selection, preparation, and training of these models.


\subsubsection{Model Selection}
\label{sec:model-selec}

State-of-the-art LLMs encompass multiple architectures (e.g., encoder-only, decoder-only, encoder-decoder), modelling paradigms (e.g., discriminative, generative), pre-training tasks (e.g., masked language modelling or MLM, permutation language modelling), size and scale, among other descriptors~\cite{zhao2023survey}. Appropriate model selection is not trivial and is often neglected or undermined. For our experiments, we opted for decoder-only models, due to their better suitability for handling classification tasks. Moreover, we avoided testing large-scale generative models due to their considerable dimensions and, therefore, their practical limitations in terms of memory and time constraints. Below we provide the selection of LLM for our research, including those features suited for our task and their role in the evaluation.

\begin{itemize}
    \item \textbf{BERT}, one of the first groundbreaking LLMs, is celebrated for its bidirectional nuanced contextual understanding~\cite{devlin2019bert}. 
    Trained on a vast corpus using MLM as pre-training objective, it excels in capturing context from both left and right, empowering it for diverse token-level tasks~\cite{broscheit-2019-investigating}. Consequently, we select BERT as a baseline for the use of LLMs in the context of feature extraction. 
    \item \textbf{RoBERTa} is considered a refinement on BERT's architecture and training process~\cite{liu2019roberta}. It achieves heightened performance through extended pre-training on a larger dataset and augmented data, resulting in more robust language representations. It also uses MLM for pre-training and outperforms BERT in various scenarios~\cite{liu2019roberta}, making it a valuable addition to our model evaluation. 
    \item \textbf{XLNet} combines autoregressive and bidirectional training by considering all possible permutations of a sentence's words during pre-training~\cite{yang2019xlnet}. This methodology fosters improved contextual understanding and dependency modelling among tokens, surpassing the conventional models. Unlike BERT and RoBERTa, XLNet uses a permutation-based training objective, allowing it to model token dependencies differently. 
\end{itemize}

For each of these models, we consider both base and large versions (i.e., in terms of number of model parameters). 

\subsubsection{Data Preparation}
\label{sec:data-prep}

We split the dataset of annotated reviews (reported in Table~\ref{tab:dataset}) under different data configurations to support different analytical perspectives.

\begin{itemize}
    \item \textbf{Out-of-domain}. The original dataset is split according to the category to which the app review belongs to. We then use these data partitions to run 10 different fine-tuning processes, using 9 out of 10 categories for training the model and using the remaining category for testing. This setup evaluates the model's capacity to generalize feature extraction to unfamiliar, new app domains. 
    \item \textbf{In-domain}. The original dataset is split under a 10-fold cross-validation setup with a balanced distribution of app reviews from each category, focusing on evaluating the model's performance when predicting features within its domain of expertise. This setup assesses the model's proficiency in feature extraction for categories closely aligned with its training data.
\end{itemize}

We exclude from all training sets all references to features included in its corresponding testing set. This allows evaluation of the model's performance to recognize tokens (extract features) for which it was not specifically fine-tuned. 

\subsubsection{Training Configuration}


For each model (Section \ref{sec:model-selec}) and data setting (Section \ref{sec:data-prep}), we configure and run a token classification fine-tuning process. First, we implement the pre-processing stage, which includes using a proper tokenizer according to the model architecture. BERT uses WordPiece tokenization and introduces [CLS] and [SEP] tokens for classification and separation. In contrast, RoBERTa and XLNet utilize SentencePiece tokenization, 
and they use only [SEP] tokens for separation while omitting the [CLS] token. Additionally, RoBERTa and XLNet employ a more aggressive subword tokenization approach, capturing finer linguistic details by breaking words into smaller subword units. This implies that a single word in the original review might be transformed into multiple tokens, which also affects the performance analysis and accuracy evaluation of the token classification (and ultimately, feature extraction) method. Second, we define the evaluation method for reporting and computing quality metrics (see Section~\ref{sec:evaluation-design}). Third, in order to adjust the experiments to the available computational resources and model characteristics, we define the training parameters for each fine-tuning process (available in the replication package). 
Finally, the output of each fine-tuning process (including checkpoints, predictions and quality metrics) for the best performing checkpoint (i.e., with the lowest evaluation loss) are saved and reported.

\subsection{Evaluation design}
\label{sec:evaluation-design}

We structure evaluation results in alignment with the formulation of research questions (Section~\ref{sec:method}). In this section, we focus on the design of the evaluation plan.

\subsubsection{Token-based ground-truth (RQ1)}
Each fine-tuning process depicted in Section~\ref{sec:data-prep} uses a token-level evaluation method for computing quality metrics for token prediction. This implies that results evaluate the model quality to predict whether a specific token is the beginning of a feature expression (\textit{B-feature}), the inner part of a feature expression (\textit{I-feature}) or none of the above (\textit{O}).     

\subsubsection{Baseline feature extraction (RQ2)}
Each fine-tuning process depicted in Section~\ref{sec:data-prep} uses a feature-level evaluation method for computing quality metrics for feature extraction. Consequently, instead of computing prediction quality at token level, in this stage we evaluate the quality prediction of complete sequences of tokens $T_f$ composing a whole feature according to the ground-truth data set. We compare the performance of our approach with respect to the baseline method selected for feature extraction (i.e., SAFE~\cite{Johann2017a}).

\begin{table*}[t!]
\centering
\caption{Token classification evaluation results.}
\vspace{-6pt} 
\begin{tabular}{@{}lll|rrrrrr@{}}
\toprule
\textbf{Analysis} &
  \textbf{Category} &
  \textbf{Metric} &
  \multicolumn{1}{c}{\textbf{BERT\textsubscript{base}}} &
  \multicolumn{1}{c}{\textbf{BERT\textsubscript{large}}} &
  \multicolumn{1}{c}{\textbf{RoBERTa\textsubscript{base}}} &
  \multicolumn{1}{c}{\textbf{RoBERTa\textsubscript{large}}} &
  \multicolumn{1}{c}{\textbf{XLNet\textsubscript{base}}} &
  \multicolumn{1}{c}{\textbf{XLNet\textsubscript{large}}} \\ \midrule
\multirow{33}{*}{\textbf{Out-of-domain}}    & \multirow{3}{*}{\textbf{PROD.}}             & \textbf{precision} & 
\cellcolor{green!25} \bestmark{0.799} & 0.734 & 0.539 & 0.287 & 0.582 & 0.687 \\
&                                                    & \textbf{recall}    & 
\bestmark{0.343} & 0.320 & 0.244 & \cellcolor{red!25}0.062 & 0.330 & 0.331 \\
&                                                    & \textbf{F1}        & 
\bestmark{0.480} & 0.445 & 0.335 & 0.102 & 0.421 & 0.447 \\
\cline{2-9}
& \multirow{3}{*}{\textbf{COMM.}}            & \textbf{precision} 
& 0.407 & \bestmark{0.502} & 0.455 & 0.384 & 0.438 & 0.412 \\
&                                                    & \textbf{recall}    & 
0.156 & 0.202 & 0.173 & 0.276 & 0.261 & \bestmark{0.317} \\
&                                                    & \textbf{F1}        & 
0.225 & 0.288 & 0.251 & 0.321 & 0.327 & \bestmark{0.358} \\
\cline{2-9} & \multirow{3}{*}{\textbf{TOOLS}}                    & \textbf{precision} &  0.513 & \bestmark{0.570} & 0.462 & 0.221 & \cellcolor{red!25}0.423 &\cellcolor{red!25} 0.214 \\
&                                                    & \textbf{recall}    &  0.085 & 0.138 & 0.102 & 0.065 & \bestmark{\cellcolor{red!25}0.204} & \cellcolor{red!25}0.026 \\
&                                                    & \textbf{F1}        &  0.145 & 0.222 & 0.167 & 0.100 & \bestmark{\cellcolor{red!25}0.275} & \cellcolor{red!25}0.046 \\
\cline{2-9}
& \multirow{3}{*}{\textbf{SOCIAL}}                   & \textbf{precision} &  0.606 & 0.696 & 0.621 & 0.610 & \bestmark{0.734} & 0.688 \\
&                                                    & \textbf{recall}    &  0.513 & 0.410 & 0.462 & 0.462 & 0.603 & \bestmark{0.679} \\
&                                                    & \textbf{F1}        &  0.556 & 0.516 & 0.529 & 0.526 & 0.662 & \bestmark{0.684} \\
\cline{2-9}
& \multirow{3}{*}{\textbf{HEALTH}}     & \textbf{precision}  & 0.482 & 0.503 & 0.658 & 0.584 & \bestmark{0.710} & 0.663 \\
&                                                    & \textbf{recall}    &  0.179 & 0.240 & 0.127 & 0.224 & 0.373 & \bestmark{0.384} \\
&                                                    & \textbf{F1}        &  0.261 & 0.325 & 0.213 & 0.323 & \bestmark{0.489} & 0.486 \\
\cline{2-9}
& \multirow{3}{*}{\textbf{PERS.}}          & \textbf{precision} & 0.731 & \cellcolor{green!25}0.955 & \cellcolor{green!25}0.933 & \cellcolor{green!25}0.973 & \cellcolor{green!25}0.972 & \bestmark{\cellcolor{green!25}1.000} \\
&                                                    & \textbf{recall}    &  0.500 & 0.553 & 0.737 & \bestmark{\cellcolor{green!25}0.947} & \cellcolor{green!25}0.921 & 0.684 \\
&                                                    & \textbf{F1}        &  0.594 & \cellcolor{green!25}0.700 & \cellcolor{green!25}0.824 & \bestmark{\cellcolor{green!25}0.960} & \cellcolor{green!25}0.946 & \cellcolor{green!25}0.813 \\
\cline{2-9}
& \multirow{3}{*}{\textbf{TRAVEL}}       & \textbf{precision}  & \bestmark{0.773} & 0.647 & 0.720 & 0.682 & 0.481 & 0.613 \\
&                                                    & \textbf{recall}    &  \cellcolor{green!25}0.708 & 0.458 & \cellcolor{green!25}0.750 & 0.625 & 0.542 & \bestmark{\cellcolor{green!25}0.792} \\
&                                                    & \textbf{F1}        &  \bestmark{\cellcolor{green!25}0.739} & 0.537 & 0.735 & 0.652 & 0.510 & 0.691 \\ 
\cline{2-9}
& \multirow{3}{*}{\textbf{MAPS}}    & \textbf{precision}  & \cellcolor{red!25}0.029 & \cellcolor{red!25}0.120 & \cellcolor{red!25}0.045 & \cellcolor{red!25}0.077 & \bestmark{0.560} & 0.467 \\
&                                                    & \textbf{recall}    &  \cellcolor{red!25}0.021 & \cellcolor{red!25}0.063 & \cellcolor{red!25}0.063 & 0.063 & \bestmark{0.292} & 0.146 \\
&                                                    & \textbf{F1}        &  \cellcolor{red!25}0.024 & \cellcolor{red!25}0.082 & \cellcolor{red!25}0.053 & \cellcolor{red!25}0.069 & \bestmark{0.384} & 0.222 \\
\cline{2-9}
& \multirow{3}{*}{\textbf{LIFESTYLE}}                & \textbf{precision} &  0.500 & 0.400 & 0.600 & 0.600 & 0.800 & \bestmark{\cellcolor{green!25}1.000} \\
&                                                    & \textbf{recall}    &  0.400 & 0.400 & 0.600 & 0.600 & \bestmark{0.800} & 0.200 \\
&                                                    & \textbf{F1}        &  0.444 & 0.400 & 0.600 & 0.600 & \bestmark{0.800} & 0.333 \\
\cline{2-9}
& \multirow{3}{*}{\textbf{WEATHER}}                  & \textbf{precision} &  0.619 & 0.642 & 0.273 & 0.129 & 0.571 & \bestmark{0.769} \\
&                                                    & \textbf{recall}    &  0.232 & \bestmark{\cellcolor{green!25}0.607} & 0.107 & 0.071 & 0.500 & 0.179 \\
&                                                    & \textbf{F1}        &  0.338 & \bestmark{0.624} & 0.154 & 0.092 & 0.533 & 0.290 \\
\cline{2-9}
& \multirow{3}{*}{\textbf{Average}}                  & \textbf{precision} &  0.546 & 0.577 & 0.531 & 0.455 & 0.627 & \bestmark{0.651} \\
&                                                    & \textbf{recall}    &  0.314 & 0.339 & 0.336 & 0.339 & \bestmark{0.482} & 0.374 \\
&                                                    & \textbf{F1}        &  0.381 & 0.414 & 0.386 & 0.374 & \bestmark{0.535} & 0.437 \\
\hline
\multirow{3}{*}{\textbf{In-domain}} & \multirow{3}{*}{\textbf{Average}} & \textbf{precision} &  0.596 & 0.719 & 0.668 & 0.688 & 0.679 & \bestmark{0.761} \\
&                                                    & \textbf{recall}    &  0.488 & \bestmark{0.582} & 0.569 & 0.509 & 0.519 & 0.573 \\
&                                                    & \textbf{F1}        &  0.532 & 0.637 & 0.611 & 0.571 & 0.582 & \bestmark{0.646} \\
\hline
\end{tabular}%
\vspace{-6pt} 
\label{tab:token-classification-results}
\end{table*}

\subsubsection{New features (RQ3)}
We select the best performing model to collect all new features predicted by our model. These features are then submitted to a human evaluation process to measure the prediction quality of new features. 
The human evaluation is composed of three main stages:

\begin{itemize}
    \item \textbf{Data preparation}. We collect all features predicted by the best-performing model (based on RQ1 and RQ2) for each test set under each data configuration scenario, as depicted in Section~\ref{sec:data-prep}. We then remove all features included in the complete ground-truth annotated dataset, keeping exclusively newly reported features.
    
    \item \textbf{Set up}. We iteratively elaborate and refine the guidelines, the selection of examples and the definition of feature annotation tasks. A task is defined as a sub set of review sentences, each one of them with a potential feature candidate which the annotator can either confirm (\textit{Yes}), reject (\textit{No}), or mark as not clear (\textit{I don't know}). 
    Figure~\ref{fig:feature-annotation-task} shows an example of a feature annotation question. 

    \begin{figure}[h]
\centerline{\includegraphics[width=0.8\columnwidth,cfbox=gray 0.2pt 1pt]{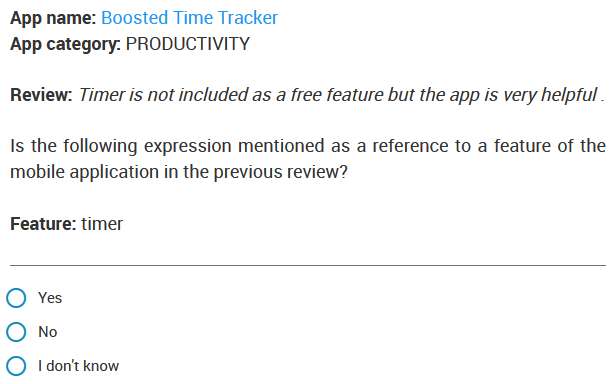}}
\caption{Example of a feature annotation question for human evaluation.}
\label{fig:feature-annotation-task}
\end{figure}
    
    This includes: app name, link to Google Play (for app context), category, review sentence, question and feature candidate.
    In this stage, we used a test task which is sequentially annotated by internal members of this research study, until an acceptable agreement is reached. After each annotation process, 
    the collected feedback is used for refining the guidelines and list of examples used for designing the evaluation task.

    \item \textbf{Evaluation}. The full data set of new features is submitted for human evaluation through sequential iterations in different batches. We used Prolific~\cite{Prolific} as a crowdsourced annotation platform to reach users worldwide and QuestBase~\cite{QuestBase} for the creation of the tasks. 
    Each annotator is limited to participate in a single task. For each task, we include a subset of 5 control questions using ground-truth annotated features to reject low-confidence annotators. On each task, we measure the proportion of features confirmed (\textit{Yes}), which relates to the precision of new features. 
    Additionally, for inter-rater reliability, we report (1) the average pairwise agreement, and (2) F1, which has been used in related work as an appropriate and effective inter-rater agreement measure for the evaluation of text annotations such as features in app reviews~\cite{Dabrowski2023}. 
\end{itemize}

\section{Evaluation}
\label{sec:evaluation}


\subsection{Token-based ground-truth}

Table~\ref{tab:token-classification-results} reports the precision, recall and F1 metrics for all data configurations and all selected models. Given that we do not have ground-truth data for non-feature entities (\textit{true negatives}), we exclude accuracy from the results.

\subsubsection{Out-of-domain Feature Extraction} In this configuration, each block in Table~\ref{tab:token-classification-results} for a given category $C$ refers to the quality metrics reported when fine-tuning the specified model with the set of reviews from all categories from Table~\ref{tab:dataset} except $C$. Metrics refer then to the test set of reviews belonging to $C$. For example, the first 3 rows in Table~\ref{tab:token-classification-results} report the performance of each model for predicting features included in app reviews from the PRODUCTIVITY when training the model with app reviews from all categories except PRODUCTIVITY. In this scenario, the best precision is reported by BERT\textsubscript{base} ($0.799$), while the lowest is reported by RoBERTa\textsubscript{large} ($0.287$). The last 3 rows in the \textit{out-of-domain} block are the average value for each metric and each model configuration. For example, the highest average recall among all categories is reported by XLNet\textsubscript{base} ($0.482$), followed by XLNet\textsubscript{large} ($0.374$). 

\begin{figure}[t]
\centerline{\includegraphics[width=\columnwidth]{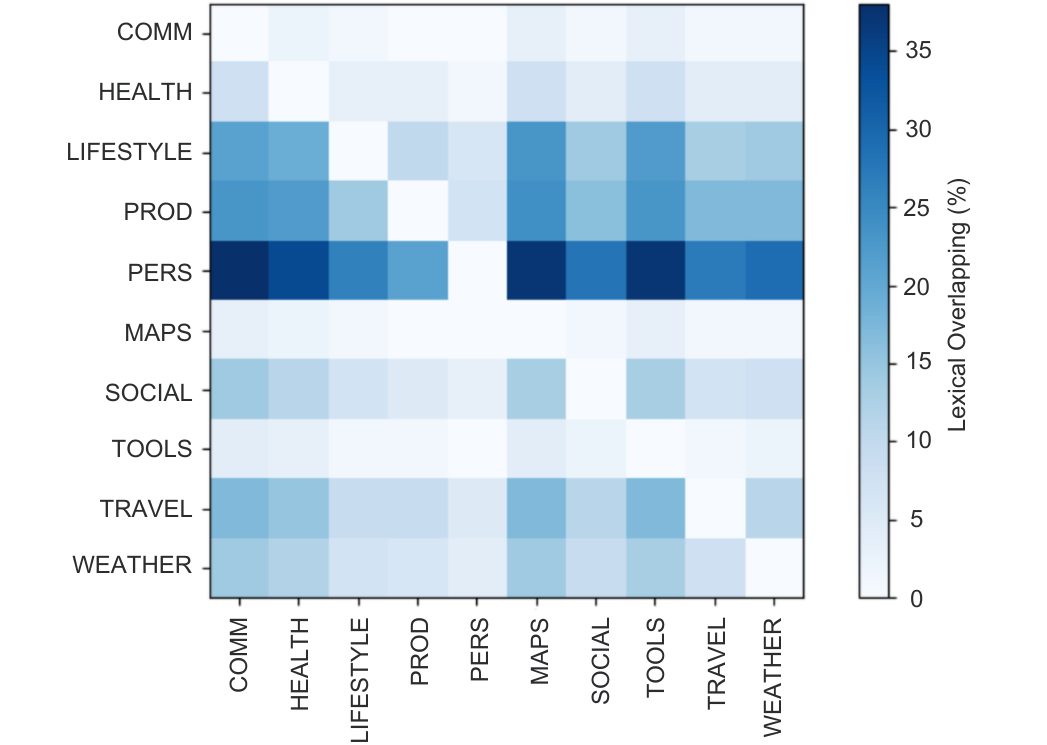}}
\caption{Lexical overlap between reviews from different categories. 
}
\vspace{-6pt} 
\label{fig:lexical-overlapping}
\vspace{-12pt} 
\end{figure}

Complementarily, we extend the visualization of the results in two dimensions. A vertical analysis illustrates the comparison between different categories for a given model configuration. We use a colour-code pattern to highlight the best (green) and worst (red) performing category for each model. For example, for the baseline model (BERT\textsubscript{base}), prediction of PRODUCTIVITY features reports the highest precision ($0.799$), while TRAVEL reports the highest recall ($0.708$), F1 ($0.739$) and accuracy ($0.966$). On the other hand, predicting new features from the MAPS domain reports the lowest overall accuracy for BERT\textsubscript{base}. If we focus on F1, predicting PERSONALIZATION features under a set-up where the model was not trained under features of this domain reports the best results for 5 out of 6 model configurations. Contrarily, predicting MAPS features under the same set-up reports the lowest overall token-level prediction quality for 4 out of 6 model configurations. For a better understanding of this phenomenon, Figure~\ref{fig:lexical-overlapping} showcases the degree (expressed in \% of tokens) of lexical overlapping of the set of reviews of a given category (Y axis) with respect to the set of reviews from another category (X axis). We exclusively considered verbs, nouns and adjectives. For example, the cell on (0,0) coordinates illustrates the proportion of tokens from WEATHER apps that are also present in reviews from the COMMUNICATION app (around 15\%). Consistently with our previous results, PERSONALIZATION presents a high overlap with all other categories (30-35\%), which showcases that the training set of reviews used for this configuration includes a high proportion of lexicon that the model has also been trained with. PERSONALIZATION apps often expose extended functions and customization capabilities to other apps, including widgets, wallpapers, stickers, themes and optimization tools. On the other hand, MAPS apps report a very low lexical overlap with respect to all categories (<~5\%). This implies that the training set of reviews used for this setup did not include any of the category-specific lexicon from the navigational domain (e.g., GPS, GPX, POI...). Consequently, out-of-domain prediction of MAPS features becomes a challenging task. Similar conclusions can be reached by observing other categories. For example, PRODUCTIVITY apps include a large sub set of apps with features present in multiple categories (e.g., calling, note-taking, file-sharing). 

In addition, a horizontal analysis in Table~\ref{tab:token-classification-results} illustrates the best-performing model for a given category. We use bold-face style and a special icon \includegraphics[height=5pt]{figures/icon.png} to highlight the best model for each category according to each metric. Additionally, we use the same strategy to report best average metrics for out-of-domain and in-domain analysis. Overall, RoBERTa models report the worst results for all metrics in almost each category, except for PERSONALIZATION apps when focusing on recall ($0.947$) and F1 ($0.960$). For BERT checkpoints, both base (PRODUCTIVITY, TRAVEL) and large (COMMUNICATION, TOOLS) report the best metrics for precision. Nevertheless, XLNet variants excel in the majority of categories, especially if we focus on recall. Specifically, on average, XLNet\textsubscript{base} reports the best recall ($0.482$) and F1 ($0.535$), while XLNet\textsubscript{large} reports the highest precision ($0.652$), but only by a small difference with respect to XLNet\textsubscript{base} ($+0.024$).

Given the limitations of the ground-truth generation, 
precision results must be interpreted under certain constraints. The lack of a guarantee of the exhaustivity of the crowdsourced features annotated by users implies that there might be tokens predicted by our model that are rejected as tokens from a feature (\textit{false positives}), while they might be part of an actual feature (\textit{true positives}) that was not indexed in the original set of features. Given that precision is the proportion of correct feature tokens with respect to all reported named entities, we argue that precision values reported above can be interpreted as the lower threshold of the minimum precision raised by our approach. On the other hand, recall (i.e., the proportion of retrieved named entities with respect to all ground-truth named entities) can be considered as a gold metric for quality analysis. All in all, for out-of-domain feature extraction, we argue that the best performing model is XLNet\textsubscript{base}. 

\begin{table*}[t]
\centering
\caption{Feature extraction evaluation results and comparison with baseline method.}
\vspace{-6pt} 
\resizebox{\textwidth}{!}{%
\begin{tabular}{@{}ll|rrrrrrrrrrr|r@{}}
\toprule
\multicolumn{1}{l}{\textbf{}}  & \textbf{}
   &
  \multicolumn{11}{c}{\textbf{Out-of-domain}} &
  \multicolumn{1}{c}{\textbf{In-domain}} \\ \hline
\multicolumn{1}{l}{\textbf{Method}} &
\multicolumn{1}{l}{\textbf{Metric}} &
  \multicolumn{1}{r}{\textbf{PROD}} &
  \multicolumn{1}{r}{\textbf{COMM}} &
  \multicolumn{1}{r}{\textbf{TOOLS}} &
  \multicolumn{1}{r}{\textbf{SOCIAL}} &
  \multicolumn{1}{r}{\textbf{HEALTH}} &
  \multicolumn{1}{r}{\textbf{PERS}} &
  \multicolumn{1}{r}{\textbf{TRAVEL}} &
  \multicolumn{1}{r}{\textbf{MAPS}} &
  \multicolumn{1}{r}{\textbf{LIFESTYLE}} &
  \multicolumn{1}{r}{\textbf{WEATHER}} &
  \multicolumn{1}{r}{\textbf{Average}} &
  \multicolumn{1}{r}{\textbf{Average}} \\ \midrule
\multirow{3}{*}{\textbf{SAFE}}      & \textbf{precision} & 0.309 & 0.235 & 0.299 & 0.286 & 0.298 & 0.195 & 0.300 & \cellcolor{red!25}0.074 &\cellcolor{green!25} 0.500 & 0.413 & 0.301 & 0.193 \\
                                    & \textbf{recall}    & \bestmark{0.300} & \bestmark{0.229} & \bestmark{0.292} & 0.329 & 0.274 & 0.229 & 0.375 & \cellcolor{red!25}0.106 & \cellcolor{green!25}0.500 & \bestmark{0.481} & 0.321 & 0.215 \\
                                    & \textbf{F1}        & \bestmark{0.304} & \bestmark{0.232} & \bestmark{0.295} & 0.306 & 0.285 & 0.211 & 0.333 & \cellcolor{red!25}0.087 & \cellcolor{green!25}0.500 & 0.444 & 0.310 & 0.199 \\ \hline
\multirow{3}{*}{\textbf{BERT\textsubscript{base}}}  & \textbf{precision} & \bestmark{0.667} & 0.278 & \bestmark{0.392} & 0.361 & 0.335 & \cellcolor{green!25}0.895 & \bestmark{0.591} & \cellcolor{red!25}0.000 & 0.500 & \bestmark{0.690} & 0.471 & 0.575 \\
                                    & \textbf{recall}    & 0.128 & 0.131 & 0.222 & 0.385 & 0.186 & 0.447 & 0.542 & \cellcolor{red!25}0.000 & \cellcolor{green!25}0.600 & 0.357 & 0.300 & 0.419 \\
                                    & \textbf{F1}        & 0.215 & 0.178 & 0.284 & 0.373 & 0.240 & \cellcolor{green!25}0.596 & \bestmark{0.565} & \cellcolor{red!25}0.000 & 0.545 & \bestmark{0.471} & 0.347 & 0.485 \\ \hline
\multirow{3}{*}{\textbf{XLNet\textsubscript{base}}} & \textbf{precision} & 0.399 & \bestmark{0.479} & 0.347 & \bestmark{0.561} & \bestmark{0.494} & \cellcolor{green!25}\bestmark{0.912} & 0.538 & \cellcolor{red!25}\bestmark{0.200} & \bestmark{0.600} & 0.502 & \bestmark{0.503} & \bestmark{0.631} \\
                                    & \textbf{recall}    & 0.244 & 0.145 & 0.168 & \bestmark{0.590} & \bestmark{0.402} & \cellcolor{green!25}\bestmark{0.816} & \bestmark{0.583} & \cellcolor{red!25}\bestmark{0.188} & \bestmark{0.600} & 0.432 & \bestmark{0.417} &  \bestmark{0.572} \\
                                    & \textbf{F1}        & 0.303 & 0.222 & 0.226 & \bestmark{0.575} & \bestmark{0.443} & \cellcolor{green!25}\bestmark{0.861} & 0.560 & \cellcolor{red!25}\bestmark{0.194} & \bestmark{0.600} & 0.464 & \bestmark{0.445} &  \bestmark{0.600}    
   \\ \hline
\end{tabular}%
}
\vspace{-12pt} 
\label{tab:baseline-tab}
\end{table*}

\subsubsection{In-domain Feature Extraction}

The last rows in Table~\ref{tab:token-classification-results} report average results for the 10-fold cross-validation analysis using the complete dataset in Table~\ref{tab:dataset}. While features in the test set for each data partition are not present in the training set (as explained in Section~\ref{sec:data-prep}), domain-related features from the same category are distributed in balance across all data splits. As expected, average results for all metrics are significantly higher in the in-domain analysis with respect to the out-of-domain analysis. 
This result indicates that language models enhance their feature extraction capabilities for a specific category $C$ when their training dataset includes reviews from that category, even if the predicted features are absent from the original training dataset. This underscores the relevance of domain-specific training data for the improvement of model performance in feature extraction tasks.

XLNet\textsubscript{large} reports the best results for precision ($0.761$) and F1 ($0.646$). In addition, XLNet\textsubscript{large} recall ($0.573$) is only slightly below ($-0.009$) with respect to BERT\textsubscript{large} recall ($0.582$). Consequently, we argue that the best performing model for in-domain feature extraction is XLNet\textsubscript{large}.

\subsection{Baseline feature extraction}
\label{sec:baseline}

Table~\ref{tab:baseline-tab} reports out-of-domain and in-domain results for the selected feature extraction baseline method (i.e., SAFE), our baseline model (i.e., BERT\textsubscript{base}) and best-performing model (i.e., XLNet\textsubscript{base}). Due to space constraints, we exclude results from other models (available in the replication package). 
Metrics in Table~\ref{tab:baseline-tab} are computed using exact matches with the whole feature (i.e., \textit{B-feature} tokens followed by none or any sequence of \textit{I-feature} tokens). 
For presentation purposes, we invert the dimensions of Table~\ref{tab:baseline-tab} with respect to Table~\ref{tab:token-classification-results}, meaning that the horizontal dimensions relate to the comparison of categories (green/red), and the vertical dimension relates to different methods (bold and \includegraphics[height=5pt]{figures/icon.png} icon).

On average, both the baseline model (BERT\textsubscript{base}) and the best-performing model (XLNet\textsubscript{base}) surpass SAFE's quality metrics for out-of-domain and in-domain analyses. The only exception is the BERT\textsubscript{base} recall ($0.300$), which is slightly below ($-0.021$) from SAFE. XLNet\textsubscript{base} improves significantly the performance with respect to BERT\textsubscript{base}, especially for recall ($+0.117$ for out-of-domain, $+0.153$ for in-domain). On a category-level, for the out-of-domain analysis, LLM-based approaches report a higher precision in all categories, and a higher recall for 6 out of 10 categories. On a horizontal analysis, similarly to results in Table~\ref{tab:token-classification-results}, predicting out-of-domain features from MAPS reviews reports the worst results in all scenarios, while the best results are provided by LIFESTYLE and PERSONALIZATION. If we focus on the categories for which SAFE outperforms T-FREX on the out-of-domain analysis, 
we realize that the original SAFE approach was designed using feature syntactic patterns from apps belonging to these domains (PRODUCTIVITY, COMMUNICATION, TOOLS). Nevertheless, as the out-of-domain configuration implies that the model was not trained using any reviews from these categories, certain limitations 
are expected. However, the in-domain analysis, which illustrates the performance of T-FREX discovering new features from a domain for which it was fine-tuned, reports a substantial improvement for precision ($+0.438$), recall ($+0.357$) and F1 ($+0.401$).

\subsection{New Features}
\label{sec:new-features}

\begin{table*}[]
\centering
\caption{Evaluation of new features}
\vspace{-6pt} 
\resizebox{\textwidth}{!}{%
\begin{tabular}{@{}l|rrrrrrrr|r|r@{}}
\toprule
\textbf{} &
  \multicolumn{8}{c}{\textbf{Evaluation \textit{(external)}}} & \multicolumn{1}{c}{\textbf{Gr. truth \textit{(external)}}} & \multicolumn{1}{c}{\textbf{Test \textit{(internal)}}} \\ \midrule
\textbf{} &
  \multicolumn{1}{r}{\textbf{PROD.}} &
  \multicolumn{1}{r}{\textbf{COMM.}} &
  \multicolumn{1}{r}{\textbf{TOOLS}} &
  \multicolumn{1}{r}{\textbf{SOCIAL}} &
  \multicolumn{1}{r}{\textbf{HEALTH}} &
  \multicolumn{1}{r}{\textbf{TRAVEL}} &
  \multicolumn{1}{r}{\textbf{MAPS}} &
  \multicolumn{1}{r}{\textbf{TOTAL}} & \multicolumn{1}{r}{\textbf{TOTAL}} & \multicolumn{1}{r}{\textbf{TOTAL}} \\ \midrule
\textbf{\#features \textit{(annotations)}} & 459    & 643    & 560    & 44     & 218    & 8      & 29     & 1956  & 190 & 95 \\
\textbf{\#features \textit{(distinct)}}    & 294    & 383    & 363    & 36     & 155    & 8      & 19     & 1067 & 144 & 17  \\ \hline
\textbf{\% Yes}                   & 68.6\% & 62.3\% & 58.4\% & 63.6\% & 59.4\% & 66.7\% & 58.6\% & 61.2\% & 77.0\% & 70.0\% \\
\textbf{\% No}                    & 28.8\% & 35.0\% & 41.7\% & 34.1\% & 39.3\% & 33.3\% & 41.4\% & 37.0\% & 23.0\% & 28.0\%\\
\textbf{\% I don't know}          & 1.6\%  & 2.7\%  & 1.8\%  & 2.2\%  & 0.6\%  & 0.0\%  & 0.0\%  & 1.9\% & 0.0\% & 2.0\% \\ \hline
\textbf{Pairwise agreement}  & 58.5\% & 58.0\% & 62.1\% & 60.2\% & 62.4\% & 57.7\% & 69.7\% & 58.5\% & 58.0\% & 73.3\% \\
\textbf{F1}                  & 0.585  & 0.566  & 0.622  & 0.594  & 0.631  & 0.639  & 0.717  & 0.613  & 0.546 & 0.719 \\ \bottomrule
\end{tabular}%
}
\vspace{-12pt} 
\label{tab:new-features}
\end{table*}

Table~\ref{tab:new-features} reports data and metrics resulted from the evaluation of new features. We discuss this results in alignment with the evaluation process depicted in Section~\ref{sec:evaluation-design}.

\subsubsection{Data preparation}
We select XLNet\textsubscript{base} as the best model for out-of-domain feature extraction (focusing on recall as the most reliable quality metric). After processing all reviews under each out-of-domain data configuration, we collected a total amount of 1,067 unique new features (1,956 annotations in total). We excluded 3 out of 10 categories from this analysis (PERSONALIZATION, LIFESTYLE, WEATHER) where either 1 or even no new features were extracted. The whole set of 1,956 review sentences was split into 21 annotation tasks. Each task included 95 new feature annotations (as in Figure~\ref{fig:feature-annotation-task}) plus 5 control questions, with 100 annotations in total per task. Additionally, we prepared 2 ground-truth tasks of the same size containing at least 1 instance of each of the distinct features available in the crowdsourced data set in Table~\ref{tab:dataset}. The purpose of these tasks is to compare the overall precision of undocumented features with respect to the overall, perceived precision from users with respect to the ground-truth.

\subsubsection{Evaluation set up} 
We conducted up to 3 iterations of internal (i.e., expert) annotations for the test task. Each iteration was performed by a different annotator, providing feedback about the guidelines, the examples, the required time to conduct the task, and the difficulty. At the end of each iteration, we refined the guidelines (focusing on reducing ambiguity), extended the examples (focusing on covering exhaustively the different kinds of features) and adjusted the expected resolution time. As a result, we decided to keep task size to 100 features and an estimated average time of 15' per task. On the test task, we report an average pairwise agreement between internal annotators of 73.3\%, and an average F1 of $0.719$. Complementarily, for internal evaluation, given that guidelines and annotators were selected and instructed under a test set-up, and all of them annotated the same features, we measured the Fleiss kappa agreement between all annotators, reporting a substantial degree of agreement ($0.718$).
    
\subsubsection{Evaluation}
We set as acceptance criteria for annotators to reply correctly to 4 out of 5 control questions, and we ran multiple iterations until reaching 5 accepted annotators for each task. We recruit participants by taking into consideration only fluent English speakers and without any language-related disorders. Each annotator is paid \$2 per task (around 15') for their participation. Category-level and aggregated results are reported in Table~\ref{tab:new-features}. The gold label for each feature was assigned using a voting-based approach between all annotators for each review sentence and feature (any ties were resolved as `\textit{I don't know}' to reduce any biases of results). For the whole dataset, 61.2\% of new features were confirmed as true features, which leads to a precision of $0.612$ for new features. Results are generally balanced across categories with minor deviations, being TOOLS the category with less accepted features (58.4\%) and PRODUCTIVITY the category with more accepted features (68.6\%). For the ground-truth validation with external annotators (including only actual features), 77.0\% of feature annotations were accepted as true features, while 23.0\% were rejected. While there is a difference of $+$15.8\% with respect to the new features, this confirms the cognitive difficulty for actual users on the formalization of a \textit{feature}.



\section{Discussion}
\label{sec:discussion}

\subsection{Research Questions}
\label{sec:rq-discussion}

Based on the evaluation results, we consolidate and report the response to each research question defined in Section~\ref{sec:method}.

\textbf{RQ1)} Table~\ref{tab:token-classification-results} provides a comprehensive empirical evaluation of the effectiveness of the T-FREX approach at token-level. Among the different model configurations, XLNet approaches (benefitting from autoregressive methods to learn bidirectional contexts) seem to provide better results on average, especially for the out-of-domain analysis.
Concerning data configurations, T-FREX proves to be significantly effective for an in-domain setting. Since mobile app markets are generally stable and the emergence of new domains is rare, the in-domain configuration is the most practical and common application for feature extraction tasks. 
Nevertheless, out-of-domain extraction also proves effective for certain domains and model configurations, particularly when the domain lexicon is not highly specialized. These results underscore T-FREX's ability to discover new features even from an unknown domain.
Consequently, the results for both in-domain and out-of-domain analyses highlight the adaptability and potential generalization of T-FREX across various settings.


\textbf{RQ2)} Table~\ref{tab:baseline-tab} presents a comprehensive empirical evaluation of the feature extraction method compared to the SAFE approach. On average, LLM-based token classification consistently outperforms SAFE across all metrics. Notably, there is a significant performance improvement when transitioning from the baseline model BERT\textsubscript{base} to XLNet\textsubscript{base}. 
The limitations of a deterministic approach, such as a non-specialized vocabulary and context-agnostic behaviour, are particularly pronounced in specific domains or categories. This holds true even when the LLMs are evaluated in an out-of-domain setting. 
Moreover, in an in-domain analysis, LLMs quickly overcome these limitations and significantly enhance their performance. They become adept at accurately predicting new features within domains they have been fine-tuned for.
In addition to leveraging pre-trained LLMs, our supervised approach can be iteratively refined and tailored to specific domains, serving specialized markets and application categories. T-FREX enables the collection of recent app reviews and the integration of up-to-date crowdsourced features, continually enhancing feature extraction through subsequent fine-tuning iterations. This approach facilitates effective context integration, adaptation to new domains, and responsiveness to users' vocabulary, syntax, and colloquial language—capabilities notably limited in SAFE and other feature extraction techniques.


\textbf{RQ3)} Table~\ref{tab:new-features} illustrates in detail the human evaluation process of new features. Results support the original hypothesis that the ground-truth dataset of features is limited. Despite these constraints, the human evaluation confirms the effectiveness of the model in the retrieval of new features. Furthermore, it is noteworthy that the precision of newly reported features surpasses the average precision of ground-truth features in the out-of-domain analysis. 
As knowledge from crowdsourced repositories is not typically exhaustive, results underscore T-FREX's ability to automatically supplement feature annotations. Instead of relying on manual user input, user feedback (i.e., reviews) can serve as a valuable resource for suggesting features in a streamlined manner, employing a voting-based mechanism for automatically extracted features. This setting can potentially address the limitations of manual annotations and reduce the imbalance in feature representation across apps with similar user interaction levels.


\subsection{Threats to Validity}
\label{sec:validity}

We assess the constraints of our study by considering the validity threats as outlined by Wohlin et al. \cite{Wohlin2014}.

Concerning construct and internal validity, we mainly relate to the formalization of \textit{features}, including its definition (used for the human evaluation in RQ3), exemplification and analysis. Related work illustrates cognitive differences in formalizing the limits of a natural language expression for a given feature (see Section~\ref{sec:back-feat-ex}). This includes the generation of the ground-truth dataset by transferring app annotations to reviews. To mitigate internal bias, we use a reliable crowdsourced software recommendation platform with real user-annotated features that are used in practice for navigation, indexing, and software comparison. Moreover, we provide a detailed analysis (see Section~\ref{sec:background}) to consolidate accepted criteria and descriptors for the formalization of a feature in the context of mobile apps.

Concerning external validity, delegating the assignment of ground-truth feature annotations to an external entity leads to a lack of control of the annotation process and the annotators (i.e, the users). 
While this entails some risks, we argue that using annotations from real users in a practical environment provides significant benefits to an LLM-based feature extraction approach. This becomes especially relevant in the context of processing reviews generated by users themselves. Additionally, RQ3 is also designed to overcome and measure the impact of missing features in the ground truth. Concerning the latter, the human evaluation process also entails external validity concerns, especially for the lack of control of the human annotators and their potential bias. To reduce this risk, we included control questions to assess the reliability of each annotation task, and we used a voting-based approach to consider the most common prediction among all annotators. 

Concerning conclusion validity, the main concern is derived from the interpretation and generalization of the performance of each evaluation setting. For this reason, we included in this study two different analytical perspectives (i.e., out-of-domain and in-domain), each including a detailed perspective on the performance of all metrics for all selected models. For the out-of-domain analysis, we also include a detailed perspective on the performance of each category. Rather than providing a gold method in a one-fits-all fashion, we aim to provide researchers with enough information to interpret the strengths and limitations of each method under each configuration. 
\section{Related work}
\label{sec:rel-work}

Dabrowski et al. recently conducted an evaluation and replication study focusing on mining techniques of app reviews for multiple tasks, including feature extraction~\cite{Dabrowski2023}. Related work mainly refers to the SAFE approach as the most consolidated technique for feature extraction~\cite{Johann2017a}. They identified and formalized a set of 18 common Part-of-Speech patterns from app descriptions and app reviews used to express app features. Through a pattern-matching approach, complemented by semantic similarity and synonymity resolution, they identify potential feature expressions. Nevertheless, reported performance in the original study is limited, especially when applying the technique to reviews, where a lot of noise features (i.e., false positives) are reported, leading to low precision. Furthermore, the original code and dataset are not publicly available. Consequently, replication studies have reimplemented and built new annotated data sets using ground-truth from instructed coders~\cite{Dabrowski2023,Shah2019}, reporting lower quality than originally reported, especially for direct feature match.

Similar conclusions apply to other related works applying the same syntactic-based strategy, either for early work~\cite{Guzman2014} or more up-to-date solutions like the ReUS approach~\cite{DRAGONI20191103}. In addition to previous limitations, evaluation strategies (including replication studies~\cite{Dabrowski2023}) are limited to instructed internal coders. The lack of the user perspective is key, especially when analysing user-generated documents (i.e., reviews). Moreover, they all focus on a reduced set of apps (8-10) and even fewer domains, and there is no formal evaluation of a category-oriented analysis for their ability to generalize to new domains. Nevertheless, they are still used in practice for feature-based knowledge generation from mobile app repositories~\cite{Kumari2022,Kasri2020}.

Few works can be found on the application of LLMs for the task of feature extraction. Similarly to previous studies~\cite{Johann2017a,Guzman2014,DRAGONI20191103}, the TransFeatEx tool applies PoS patterns by leveraging the knowledge embedded in a RoBERTa model to extract syntactic and semantic annotations~\cite{Gallego2023}. 
Nevertheless, their contribution is presented as a tool without further evaluation or a concrete proposal for configuring the pattern template or the sentiment analysis thresholds. KEFE~\cite{Wu2021922} uses features extracted using PoS patterns as input to a BERT model for text classification of correct and incorrect features. However, they focus on the application of this technique for app descriptions, using the extracted features to transfer potential feature matches with user reviews. Consequently, feature knowledge is limited to developer-generated documentation. 
\section{Conclusions and future work}
\label{sec:conclusions}

In this research, we conducted an empirical evaluation of a token classification-based approach using LLMs to support feature extraction in the context of mobile app reviews. We explored and discussed in detail the performance of multiple models (BERT, RoBERTa, XLNet) under different data configurations (out-of-domain vs. in-domain) from multiple app categories. The evaluation provides a comprehensive perspective of the performance of each approach under each data configuration. Furthermore, ground-truth feature annotations by real users and external human evaluation contribute to extending the scope and body of knowledge of the feature landscape. All in all, our proposal leverages the potential of LLMs to benefit from contextualized knowledge and to overcome the limitations of syntactic-based approaches. Research and industrial applications focusing on software evolution analysis can benefit from the outcomes of this study, either by replicating T-FREX as a fully automatic process for feature extraction (either with different data sets or different models), by using the ground-truth data set of annotated reviews, or by using any of the fine-tuned models distributed for replication. 

As future work, we are currently working on the potential of extending the pre-training of the LLMs used for evaluation with a large data set of app reviews. Evaluation will focus on token classification and feature extraction metrics with respect to the original models. Furthermore, we plan to gain a better understanding of the inner workings of these models by analysing the embedded knowledge across multiple layers. To this end, hidden layers can be used as input for probing classifiers to determine to what extent the given layer embeds relevant knowledge to support feature extraction.
\section*{Acknowledgments}

\noindent \small{With the support from the Secretariat for Universities and Research of the Ministry of Business and Knowledge of the Government of Catalonia and the European Social Fund.
This paper has been funded by the Spanish Ministerio de Ciencia e Innovación under project / funding scheme PID2020-117191RB-I00 / AEI/10.13039/501100011033.
Alessio Miaschi and Felice Dell'Orletta have also been supported by the PNRR project FAIR - Future AI Research (PE00000013), under the NRRP MUR program funded by the NextGenerationEU.}

\bibliographystyle{IEEEtran}
\bibliography{IEEEabrv,base}

\begin{thebibliography}{10}
\providecommand{\url}[1]{#1}
\csname url@samestyle\endcsname
\providecommand{\newblock}{\relax}
\providecommand{\bibinfo}[2]{#2}
\providecommand{\BIBentrySTDinterwordspacing}{\spaceskip=0pt\relax}
\providecommand{\BIBentryALTinterwordstretchfactor}{4}
\providecommand{\BIBentryALTinterwordspacing}{\spaceskip=\fontdimen2\font plus
\BIBentryALTinterwordstretchfactor\fontdimen3\font minus
  \fontdimen4\font\relax}
\providecommand{\BIBforeignlanguage}[2]{{%
\expandafter\ifx\csname l@#1\endcsname\relax
\typeout{** WARNING: IEEEtran.bst: No hyphenation pattern has been}%
\typeout{** loaded for the language `#1'. Using the pattern for}%
\typeout{** the default language instead.}%
\else
\language=\csname l@#1\endcsname
\fi
#2}}
\providecommand{\BIBdecl}{\relax}
\BIBdecl

\bibitem{Martin2017}
W.~Martin, F.~Sarro, Y.~Jia, Y.~Zhang, and M.~Harman, ``A survey of app store
  analysis for software engineering,'' \emph{IEEE Transactions on Software
  Engineering}, vol.~43, no.~9, pp. 817--847, 2017.

\bibitem{ACMNL}
\BIBentryALTinterwordspacing
{Authority for Consumers \& Markets}, ``Market study into mobile app stores
  {(Report ACM/18/032693)},'' april 2019. [Online]. Available:
  \url{https://www.acm.nl/sites/default/files/documents/market-study-into-mobile-app-stores.pdf}
\BIBentrySTDinterwordspacing

\bibitem{Maalej2015}
W.~Maalej and H.~Nabil, ``Bug report, feature request, or simply praise? on
  automatically classifying app reviews,'' in \emph{2015 IEEE 23rd
  International Requirements Engineering Conference (RE)}, 2015, pp. 116--125.

\bibitem{Hassan2022}
S.~Hassan, H.~Li, and A.~E. Hassan, ``On the importance of performing app
  analysis within peer groups,'' in \emph{2022 IEEE International Conference on
  Software Analysis, Evolution and Reengineering (SANER)}, 2022, pp. 890--901.

\bibitem{Yadav2020}
A.~Yadav, R.~Sharma, and F.~H. Fard, ``A semantic-based framework for analyzing
  app users' feedback,'' in \emph{2020 IEEE 27th International Conference on
  Software Analysis, Evolution and Reengineering (SANER)}, 2020, pp. 572--576.

\bibitem{Guerrouj2015}
L.~Guerrouj, S.~Azad, and P.~C. Rigby, ``The influence of app churn on app
  success and stackoverflow discussions,'' in \emph{2015 IEEE 22nd
  International Conference on Software Analysis, Evolution, and Reengineering
  (SANER)}, 2015, pp. 321--330.

\bibitem{Panichella2015}
S.~Panichella, A.~Di~Sorbo, E.~Guzman, C.~A. Visaggio, G.~Canfora, and H.~C.
  Gall, ``How can i improve my app? classifying user reviews for software
  maintenance and evolution,'' in \emph{2015 IEEE International Conference on
  Software Maintenance and Evolution (ICSME)}, 2015, pp. 281--290.

\bibitem{Dabrowski2022}
J.~D\k{a}browski, E.~Letier, A.~Perini, and A.~Susi, ``Analysing app reviews
  for software engineering: A systematic literature review,'' \emph{Empirical
  Softw. Engg.}, vol.~27, no.~2, mar 2022.

\bibitem{Begel2014}
A.~Begel and T.~Zimmermann, ``Analyze this! 145 questions for data scientists
  in software engineering,'' in \emph{Proceedings of the 36th International
  Conference on Software Engineering}, 2014, p. 12–23.

\bibitem{Buse2012}
R.~P.~L. Buse and T.~Zimmermann, ``Information needs for software development
  analytics,'' in \emph{2012 34th International Conference on Software
  Engineering (ICSE)}, 2012, pp. 987--996.

\bibitem{Dabrowski2023}
\BIBentryALTinterwordspacing
J.~D{\c{a}}browski, E.~Letier, A.~Perini, and A.~Susi, ``{Mining and searching
  app reviews for requirements engineering: Evaluation and replication
  studies},'' \emph{Information Systems}, vol. 114, p. 102181, 2023. [Online].
  Available: \url{https://doi.org/10.1016/j.is.2023.102181}
\BIBentrySTDinterwordspacing

\bibitem{Palomba2015}
F.~Palomba, M.~Linares-Vásquez, G.~Bavota, R.~Oliveto, M.~Di~Penta,
  D.~Poshyvanyk, and A.~De~Lucia, ``User reviews matter! tracking crowdsourced
  reviews to support evolution of successful apps,'' in \emph{2015 IEEE
  International Conference on Software Maintenance and Evolution (ICSME)},
  2015, pp. 291--300.

\bibitem{Guzman2014}
E.~Guzman and W.~Maalej, ``{How do users like this feature? A fine grained
  sentiment analysis of App reviews},'' \emph{2014 IEEE 22nd International
  Requirements Engineering Conference, RE 2014 - Proceedings}, pp. 153--162,
  2014.

\bibitem{Pagano2013}
D.~Pagano and W.~Maalej, ``User feedback in the appstore: An empirical study,''
  in \emph{2013 21st IEEE International Requirements Engineering Conference
  (RE)}, 2013, pp. 125--134.

\bibitem{Gao2018}
C.~Gao, J.~Zeng, M.~R. Lyu, and I.~King, ``Online app review analysis for
  identifying emerging issues,'' in \emph{Proceedings of the 40th International
  Conference on Software Engineering}, 2018, p. 48–58.

\bibitem{Chen2014}
N.~Chen, J.~Lin, S.~C.~H. Hoi, X.~Xiao, and B.~Zhang, ``Ar-miner: Mining
  informative reviews for developers from mobile app marketplace,'' in
  \emph{Proceedings of the 36th International Conference on Software
  Engineering}.\hskip 1em plus 0.5em minus 0.4em\relax New York, NY, USA:
  Association for Computing Machinery, 2014, p. 767–778.

\bibitem{Shah2019a}
{Shah, Faiz Ali and Sirts, Kairit and Pfahl, Dietmar}, ``{Is the SAFE Approach
  Too Simple for App Feature Extraction? A Replication Study},'' in
  \emph{{Requirements Engineering: Foundation for Software Quality}}, {Knauss,
  Eric and Goedicke, Michael}, Ed.\hskip 1em plus 0.5em minus 0.4em\relax
  {Cham}: {Springer International Publishing}, {2019}, pp. {21--36}.

\bibitem{vaswani2017attention}
A.~Vaswani, N.~Shazeer, N.~Parmar, J.~Uszkoreit, L.~Jones, A.~N. Gomez,
  {\L}.~Kaiser, and I.~Polosukhin, ``Attention is all you need,''
  \emph{Advances in neural information processing systems}, vol.~30, 2017.

\bibitem{Zhang2020}
T.~Zhang, B.~Xu, F.~Thung, S.~A. Haryono, D.~Lo, and L.~Jiang, ``Sentiment
  analysis for software engineering: How far can pre-trained transformer models
  go?'' in \emph{2020 IEEE International Conference on Software Maintenance and
  Evolution (ICSME)}, 2020, pp. 70--80.

\bibitem{Ciborowska2022}
\BIBentryALTinterwordspacing
A.~Ciborowska and K.~Damevski, ``Fast changeset-based bug localization with
  bert,'' in \emph{Proceedings of the 44th International Conference on Software
  Engineering}.\hskip 1em plus 0.5em minus 0.4em\relax New York, NY, USA:
  Association for Computing Machinery, 2022, p. 946–957. [Online]. Available:
  \url{https://doi-org.recursos.biblioteca.upc.edu/10.1145/3510003.3510042}
\BIBentrySTDinterwordspacing

\bibitem{Yang2022}
C.~Yang, B.~Xu, J.~Khan, G.~Uddin, D.~Han, Z.~Yang, and D.~Lo, ``Aspect-based
  api review classification: How far can pre-trained transformer model go?'' in
  \emph{2022 IEEE International Conference on Software Analysis, Evolution and
  Reengineering (SANER)}, 2022, pp. 385--395.

\bibitem{Tabassum2020}
\BIBentryALTinterwordspacing
J.~Tabassum, M.~Maddela, W.~Xu, and A.~Ritter, ``Code and named entity
  recognition in {S}tack{O}verflow,'' in \emph{Proceedings of the 58th Annual
  Meeting of the Association for Computational Linguistics}.\hskip 1em plus
  0.5em minus 0.4em\relax Online: Association for Computational Linguistics,
  jul 2020, pp. 4913--4926. [Online]. Available:
  \url{https://aclanthology.org/2020.acl-main.443}
\BIBentrySTDinterwordspacing

\bibitem{Jha2019}
\BIBentryALTinterwordspacing
N.~Jha and A.~Mahmoud, ``Mining non-functional requirements from app store
  reviews,'' \emph{Empirical Software Engineering}, pp. 1--37, 2019. [Online].
  Available: \url{https://api.semanticscholar.org/CorpusID:174802984}
\BIBentrySTDinterwordspacing

\bibitem{kang1990feature}
K.~C. Kang, S.~G. Cohen, J.~A. Hess, W.~E. Novak, and A.~S. Peterson,
  ``Feature-oriented domain analysis feasibility study,'' SEI Technical Report
  CMU/SEI-90-TR-21, Tech. Rep., 1990.

\bibitem{Wiegers2013}
K.~E. Wiegers and J.~Beatty, \emph{Software Requirements 3}.\hskip 1em plus
  0.5em minus 0.4em\relax USA: Microsoft Press, 2013.

\bibitem{Harman2012}
M.~Harman, Y.~Jia, and Y.~Zhang, ``App store mining and analysis: Msr for app
  stores,'' in \emph{2012 9th IEEE Working Conference on Mining Software
  Repositories (MSR)}, 2012, pp. 108--111.

\bibitem{KangFeatureOrientedDomain1990}
\BIBentryALTinterwordspacing
K.~Kang, S.~Cohen, J.~Hess, W.~Novak, and A.~Peterson, ``Feature-oriented
  domain analysis (foda) feasibility study,'' Software Engineering Institute,
  Carnegie Mellon University, Pittsburgh, PA, Tech. Rep. CMU/SEI-90-TR-021,
  1990. [Online]. Available:
  \url{http://resources.sei.cmu.edu/library/asset-view.cfm?AssetID=11231}
\BIBentrySTDinterwordspacing

\bibitem{jehangir2023survey}
B.~Jehangir, S.~Radhakrishnan, and R.~Agarwal, ``A survey on named entity
  recognition—datasets, tools, and methodologies,'' \emph{Natural Language
  Processing Journal}, vol.~3, p. 100017, 2023.

\bibitem{hakala-pyysalo-2019-biomedical}
\BIBentryALTinterwordspacing
K.~Hakala and S.~Pyysalo, ``Biomedical named entity recognition with
  multilingual {BERT},'' in \emph{Proceedings of the 5th Workshop on BioNLP
  Open Shared Tasks}.\hskip 1em plus 0.5em minus 0.4em\relax Hong Kong, China:
  Association for Computational Linguistics, Nov. 2019, pp. 56--61. [Online].
  Available: \url{https://aclanthology.org/D19-5709}
\BIBentrySTDinterwordspacing

\bibitem{Tarcar2019HealthcareNM}
\BIBentryALTinterwordspacing
A.~K. Tarcar, A.~Tiwari, D.~Rao, V.~N. Dhaimodker, P.~Rebelo, and R.~Desai,
  ``Healthcare ner models using language model pretraining,'' in
  \emph{HSDM@WSDM}, 2019. [Online]. Available:
  \url{https://api.semanticscholar.org/CorpusID:210943047}
\BIBentrySTDinterwordspacing

\bibitem{gu2020named}
L.~Gu, W.~Zhang, Y.~Wang, B.~Li, and S.~Mao, ``Named entity recognition in
  judicial field based on bert-bilstm-crf model,'' in \emph{2020 International
  Workshop on Electronic Communication and Artificial Intelligence
  (IWECAI)}.\hskip 1em plus 0.5em minus 0.4em\relax IEEE, 2020, pp. 170--174.

\bibitem{Stol2018}
\BIBentryALTinterwordspacing
K.-J. Stol and B.~Fitzgerald, ``The abc of software engineering research,''
  \emph{ACM Trans. Softw. Eng. Methodol.}, vol.~27, no.~3, sep 2018. [Online].
  Available: \url{https://doi.org/10.1145/3241743}
\BIBentrySTDinterwordspacing

\bibitem{Motger2023}
\BIBentryALTinterwordspacing
Q.~Motger, X.~Franch, and J.~Marco, ``Mobile feature-oriented knowledge base
  generation using knowledge graphs,'' in \emph{New Trends in Database and
  Information Systems - {ADBIS} 2023 Short Papers, Doctoral Consortium and
  Workshops: AIDMA, DOING, K-Gals, MADEISD, PeRS, Barcelona, Spain, September
  4-7, 2023, Proceedings}, ser. Communications in Computer and Information
  Science, vol. 1850.\hskip 1em plus 0.5em minus 0.4em\relax Springer, 2023,
  pp. 269--279. [Online]. Available:
  \url{https://doi.org/10.1007/978-3-031-42941-5\_24}
\BIBentrySTDinterwordspacing

\bibitem{Johann2017a}
T.~Johann, C.~Stanik, A.~M. Alizadeh, and W.~Maalej, ``{SAFE: A Simple Approach
  for Feature Extraction from App Descriptions and App Reviews},''
  \emph{Proceedings - 2017 IEEE 25th International Requirements Engineering
  Conference, RE 2017}, pp. 21--30, 2017.

\bibitem{DRAGONI20191103}
M.~Dragoni, M.~Federici, and A.~Rexha, ``An unsupervised aspect extraction
  strategy for monitoring real-time reviews stream,'' \emph{Information
  Processing \& Management}, vol.~56, no.~3, pp. 1103--1118, 2019.

\bibitem{GooglePlayCategories}
\BIBentryALTinterwordspacing
AppTweak, ``{Google Play Store Categories},'' 2022, {Accessed 5th October,
  2023}. [Online]. Available:
  \url{https://developers.apptweak.com/reference/google-play-store-categories}
\BIBentrySTDinterwordspacing

\bibitem{Stanza}
\BIBentryALTinterwordspacing
{\relax Stanford NLP Group}, ``Neural pipeline.'' [Online]. Available:
  \url{https://stanfordnlp.github.io/stanza/neural\_pipeline.html}
\BIBentrySTDinterwordspacing

\bibitem{CoNLL}
\BIBentryALTinterwordspacing
{\relax Universal Dependencies}, ``{CoNLL-U Format}.'' [Online]. Available:
  \url{https://universaldependencies.org/format.html}
\BIBentrySTDinterwordspacing

\bibitem{zhao2023survey}
W.~X. Zhao, K.~Zhou, J.~Li, T.~Tang, X.~Wang, Y.~Hou, Y.~Min, B.~Zhang,
  J.~Zhang, Z.~Dong \emph{et~al.}, ``A survey of large language models,''
  \emph{arXiv preprint arXiv:2303.18223}, 2023.

\bibitem{devlin2019bert}
\BIBentryALTinterwordspacing
J.~Devlin, M.-W. Chang, K.~Lee, and K.~Toutanova, ``{BERT}: Pre-training of
  deep bidirectional transformers for language understanding,'' in
  \emph{Proceedings of the 2019 Conference of the North {A}merican Chapter of
  the Association for Computational Linguistics: Human Language Technologies,
  Volume 1 (Long and Short Papers)}.\hskip 1em plus 0.5em minus 0.4em\relax
  Minneapolis, Minnesota: Association for Computational Linguistics, Jun. 2019,
  pp. 4171--4186. [Online]. Available: \url{https://aclanthology.org/N19-1423}
\BIBentrySTDinterwordspacing

\bibitem{broscheit-2019-investigating}
\BIBentryALTinterwordspacing
S.~Broscheit, ``Investigating entity knowledge in {BERT} with simple neural
  end-to-end entity linking,'' in \emph{Proceedings of the 23rd Conference on
  Computational Natural Language Learning (CoNLL)}.\hskip 1em plus 0.5em minus
  0.4em\relax Hong Kong, China: Association for Computational Linguistics, Nov.
  2019, pp. 677--685. [Online]. Available:
  \url{https://aclanthology.org/K19-1063}
\BIBentrySTDinterwordspacing

\bibitem{liu2019roberta}
Y.~Liu, M.~Ott, N.~Goyal, J.~Du, M.~Joshi, D.~Chen, O.~Levy, M.~Lewis,
  L.~Zettlemoyer, and V.~Stoyanov, ``Roberta: A robustly optimized bert
  pretraining approach,'' 2019.

\bibitem{yang2019xlnet}
Z.~Yang, Z.~Dai, Y.~Yang, J.~Carbonell, R.~R. Salakhutdinov, and Q.~V. Le,
  ``Xlnet: Generalized autoregressive pretraining for language understanding,''
  \emph{Advances in neural information processing systems}, vol.~32, 2019.

\bibitem{Prolific}
\BIBentryALTinterwordspacing
``{Prolific · Quickly find research participants you can trust.}'' [Online].
  Available: \url{https://www.prolific.com/}
\BIBentrySTDinterwordspacing

\bibitem{QuestBase}
\BIBentryALTinterwordspacing
``{QuestBase}.'' [Online]. Available:
  \url{https://questbase.com/en/home-questbase/}
\BIBentrySTDinterwordspacing

\bibitem{Wohlin2014}
C.~Wohlin, ``Guidelines for snowballing in systematic literature studies and a
  replication in software engineering,'' in \emph{Proceedings of the 18th
  International Conference on Evaluation and Assessment in Software
  Engineering}, 2014.

\bibitem{Shah2019}
F.~A. Shah, K.~Sirts, and D.~Pfahl, ``Using app reviews for competitive
  analysis: Tool support,'' in \emph{MAWA 2019}, 2019, pp. 40--46.

\bibitem{Kumari2022}
S.~Kumari and Z.~A. Memon, ``Extracting feature requests from online reviews of
  travel industry,'' \emph{Acta Scientiarum - Technology}, vol.~44, 2022.

\bibitem{Kasri2020}
M.~Kasri \emph{et~al.}, ``{A Comparison of Features Extraction Methods for
  Arabic Sentiment Analysis},'' in \emph{4th International Conference on Big
  Data and Internet of Things}, 2020.

\bibitem{Gallego2023}
\BIBentryALTinterwordspacing
A.~Gállego, J.~Motger, X.~Franch, and J.~Marco, ``{TransFeatEx: a NLP pipeline
  for feature extraction},'' in \emph{Joint proceedings of REFSQ-2023
  Workshops, Doctoral Symposium, Posters \& Tools Track and Journal Early
  Feedback: co-located with the 28th International Conference on Requirements
  Engineering: Foundation for Software Quality (REFSQ 2023): Barcelona,
  Catalunya, Spain, April 17-20, 2023}.\hskip 1em plus 0.5em minus 0.4em\relax
  CEUR-WS.org, 2023. [Online]. Available:
  \url{https://ceur-ws.org/Vol-3378/PT-paper1.pdf}
\BIBentrySTDinterwordspacing

\bibitem{Wu2021922}
H.~Wu \emph{et~al.}, ``{Identifying key features from app user reviews},'' in
  \emph{International Conference on Software Engineering}, 2021.

\end{thebibliography}

\end{document}